\documentclass{emulateapj} 
\shorttitle{Shell-shocked model for SN2006gy} 
\shortauthors{Smith et al.} 
\newcommand\td{$\tau_{\tighten diff}$} 
\newcommand\te{$\tau_{\tighten exp}$} 
\begin{document}

\title{Shell-shocked diffusion model for the light curve of SN~2006\lowercase{gy}} 

\author{Nathan Smith\altaffilmark{1} and Richard McCray\altaffilmark{2}} 

\altaffiltext{1}{Astronomy Department, University of California, 601
Campbell Hall, Berkeley, CA 94720}

\altaffiltext{2}{JILA, University of Colorado, 440 UCB, Boulder, CO
80309}

\begin{abstract} 

We explore a simple model for the high luminosity of SN~2006gy
involving photon diffusion of shock-deposited thermal energy. The
distinguishing property of the model is that the large ``stellar''
radius of $\sim$160 AU required to prevent adiabatic losses is not the
true stellar radius, but rather, the radius of an opaque, unbound
circumstellar envelope, created when $\sim$10 M$_{\odot}$ was ejected
in the decade before the supernova in an eruption analogous to that of
$\eta$~Carinae.  The supernova light is produced primarily by
diffusion of thermal energy following the passage of the blast wave
through this shell.  This model differs from traditional models of
supernova debris interacting with external circumstellar matter (CSM)
in that here the shell is optically thick and the escape of radiation
is delayed.  We show that any model attempting to account for
SN~2006gy's huge luminosity with radiation emitted by ongoing CSM
interaction fails for the following basic reason: the CSM density
required to achieve the observed luminosity makes the same
circumstellar envelope opaque ($\tau\ga$300), forcing a thermal
diffusion solution.  In our model, the weaker CSM interaction giving
rise to SN~2006gy's characteristic Type IIn spectrum and soft X-rays
is not linked to the power source of the visual continuum; instead, it
arises {\it after} the blast wave breaks free of the opaque shell into
the surrounding wind.  While a simple diffusion model can explain the
gross properties of the early light curve of SN~2006gy, it predicts
that the light curve must plummet rapidly at late-times, unless an
additional power source is present.

\end{abstract} 


\section{INTRODUCTION} 

The recent extremely luminous supernova (SN) 2006gy has challenged our
understanding of the deaths of massive stars and may represent a new
class of explosions. It was discovered by Quimby (2006) and was
subsequently studied in detail by Ofek et al.\ (2007) and Smith et
al.\ (2007). It is a Type IIn event, with strong narrow lines of H in
its spectrum, although the details of its light curve and spectrum are
unlike most other members of this class (e.g., Filippenko 1997). Its
enormous total radiated energy of $>$10$^{51}$ ergs, the late
($\sim$70 days) peak of the light curve, and the observed constant
expansion speed of $v_{exp}\simeq$4000 km s$^{-1}$ (Smith et al.\
2007) place important constraints on the model.  SN~2006gy exploded in
the host galaxy NGC~1260, which is a peculiar S0/Sa galaxy with
ongoing star formation (see Smith et al.\ 2007 and Ofek et al.\ 2007).

The optical displays of all Type Ia SNe and most Type II SNe are
believed to be dominated by the radioactive decay sequence $^{56}$Ni$
\rightarrow ^{56}$Co$ \rightarrow ^{56}$Fe.  If that is also the case
for SN~2006gy, a mass M($^{56}$Ni)$\approx$10 M$_{\odot}$ is required
to produce the observed luminosity (Smith et al.\ 2007).
 
Not surprisingly, SN~2006gy has prompted varying suggestions to
account for its extreme energy budget. Ofek et al.\ (2007) proposed a
SN~Ia exploding in a dense H-rich circumstellar medium (CSM) where
continual interaction with an external CSM was the power source for
the luminosity.  They suggested that the dense CSM was probably the
result of binary star common envelope ejection, and that the extreme
energy budget may need to draw upon a super-Chandrasekhar Type~Ia
explosion or perhaps a massive star explosion. Smith et al.\ (2007)
argued that many different spectral properties of SN~2006gy and its
CSM were incompatible with a Type~Ia explosion, but were entirely
consistent with the known properties of some very massive stars.  To
account for the luminosity with CSM interaction, Smith et al.\ (2007)
proposed that the star suffered a tremendous explosive but
non-terminal mass-loss event in the decade preceding the SN, analogous
to the 19th century eruption of $\eta$~Carinae. Woosley et al.\ (2007)
proposed a similar model, where the $\eta$~Car-like explosion in the
decade preceding the SN could have been triggered by the pulsational
pair instability (different from a genuine pair instability SN) in a
star with initial mass 110 M$_{\odot}$.

Interaction with an external CSM can, in principle, generate high
luminosity as seen in other Type IIn SNe (e.g., Chugai et al.\
2004). However, Smith et al.\ (2007) argued that this mechanism is
difficult to reconcile with the relatively low progenitor mass-loss
rate inferred from the weak X-rays detected by {\it Chandra} and the
narrow H$\alpha$ emission in SN~2006gy's spectrum. In other words,
signs of CSM interaction are seen in SN~2006gy, but that interaction
appears far too weak to power the visual continuum.  Even more
troubling, the CSM interaction hypothesis cannot account for why the
blast wave apparently did not decelerate while it was drained of more
than 10$^{51}$ ergs (Smith et al.\ 2007). These inconsistencies,
combined with the slow rise, low expansion speed, and high luminosity
of SN~2006gy, led Smith et al.\ (2007) to appeal to the alternative of
a pair-instability SN (Barkat et al.\ 1967; Bond et al.\ 1984; Heger
\& Woosley 2002) where the core of the star is obliterated and the
light is powered by radioactive decay of $\sim$10 M$_{\odot}$ of
$^{56}$Ni. As we discuss below, however, it may be possible to account
for the luminosity of SN~2006gy without placing such extreme demands
on SN nucleosynthesis.

The most pressing question about SN~2006gy boils down to this: Do we
{\it need} a pair instability SN to provide a consistent explanation
for its extremely high luminosity and spectral properties?  The
pair-instability hypothesis requires the introduction of an exotic
phenomenon; while the idea has been around for decades, it has never
been observed and is only expected to have occurred in the early
Universe (e.g., Heger \& Woosley 2002). Published models of the light
curves from pair instability SNe do indeed predict high luminosities,
slow rise times, long durations, and slow expansion speeds (e.g.,
Scannapieco et al.\ 2005) qualitatively similar to SN~2006gy. The
light curve shape depends on the assumed mass of the envelope, and
most studies so far have been conducted for zero-metallicity stars
with no mass loss. Further work is in progress, with various
assumptions about the mass of the envelope (Nomoto et al.\ 2007; Kasen
2007; Young 2007). The progenitor of SN~2006gy is likely to have been
a very massive star (Smith et al.\ 2007), but whether it was massive
enough to suffer the pair instability remains an open question.

Here we describe how a simple photon diffusion model can account for
the high luminosity and long duration of SN~2006gy.  Since the
explosions of core collapse SNe typically deposit $E_0 \simeq 10^{51}$
ergs of energy in the SN envelope, an exceptional mass of $^{56}$Ni is
not necessarily required for SN~2006gy if an efficient mechanism could
convert the energy of the initial SN blast into emergent light.  In
SNe from compact progenitors, the light from the blast wave itself
normally falls short by orders of magnitude.  The trapped internal
radiation will cool by virtue of adiabatic expansion, and its energy
will consequently be tranferred to the kinetic energy of the expanding
debris.  The fraction of the blast energy emerging as radiation will
be roughly $R_0/R_{max}$, where $R_0$ is the initial radius of the
progenitor and $R_{max}$ is the radius of the SN photosphere at peak
light. $R_0$ ranges from $\sim$10$^{-2}$ AU for the Wolf-Rayet
progenitors of SNe Ib/c to $\sim$1 AU for the red supergiant
progenitors of SNe~II, while typically, $R_{max} \simeq 100$ AU.

Falk \& Arnett (1973, 1977) showed that a SN from a progenitor having
a very extended envelope could produce a light curve with a long
timescale and high luminosity, with very efficient conversion of blast
wave energy into radiation.  Below we consider this idea as a possible
explanation for the light curve of SN2006gy. While we argue that this
mechanism may provide a plausible explanation for the early light
curve, we caution that it {\it does not negate} the possibility that
SN~2006gy is powered by some other mechanism at late times, such as
$^{56}$Co decay or continued CSM interaction.
 
 
\section{SHELL-SHOCKED MODEL} 

We have in mind a model in which the SN progenitor (having mass
$M_{SN}$ and radius $\la$1 A.U.) is enshrouded by an opaque shell
having mass $M_{shell}$ and characteristic radius $R_0\simeq$160 A.U.
The SN blast imparts $E_0 = 10^{51} E_{51}$ ergs of kinetic energy.
When the SN debris strike the shell, roughly $E_0 \ M_{shell}/2(M_{SN}
+ M_{shell})$ of the total SN energy will be converted into radiation
by the resulting forward and reverse shocks.  This radiation will be
produced initially as X-rays, but will quickly thermalize to an
approximate Planck spectrum with substantially lower temperature
determined by the post-shock thermal energy density.  The radiation
will escape with little adiabatic loss if the characteristic diffusion
timescale, \td, is comparable to or less than the expansion timescale,
\te, of the ejecta.  If the opacity is dominated by electron
scattering, the characteristic time for photon diffusion is \
\td$\simeq$($n \ \sigma_T \ R^2$)/$c$, where $n$ is the electron
density, $\sigma_T$=6.65$\times$10$^{-25}$ cm$^{2}$ is the Thomson
scattering cross section, $R$ is the initial radius, and $c$ is the
speed of light. For a uniform density spherical envelope with Solar
composition, this simplifies to roughly \ \td$\simeq$ 23 days
($M/R_{15}$), where $M$ is the mass of the stellar envelope in
M$_{\odot}$ and $R_{15}$ is its radius in units of 10$^{15}$ cm. The
expansion timescale is just \te$\simeq$($\Delta$R/$v_{exp}$) for
constant expansion speed.
 
In a thermal diffusion model for SN~2006gy, we take the light curve
peak at 70 days (Smith et al.\ 2007) as representative of both \td \
and \te.  From the observed constant expansion speed of
$v_{exp}\simeq$4000 km s$^{-1}$ (Smith et al.\ 2007), we have $\Delta
R \simeq v_{exp}$\te = 2.4$\times$10$^{15}$ cm or about 160 AU. Thus,
for \td = \te, the mass of ejecta through which the radiation must
travel is $\sim$10 M$_{\odot}$.
 
\begin{figure} 
 \epsscale{1.2} 
 \plotone{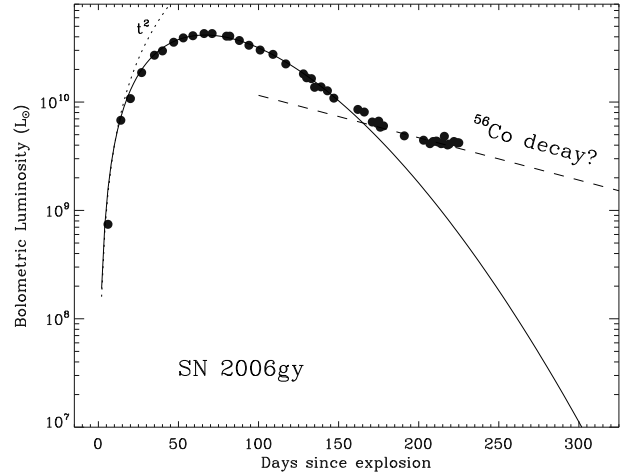} 
 \caption{The light curve of SN2006gy from Smith et al.\ (2007)
 converted to Solar luminosities from $M_R$, assuming zero bolometric
 correction. The solid curve shows a simple light curve for a model of
 thermal diffusion from an extended stellar envelope (see text). The
 dashed line shows the hypothetical $^{56}$Co decay luminosity
 expected for 8 M$_{\odot}$ of nickel.}
\end{figure} 
 
The visual light curve of SN~2006gy is unlike any other SN~IIn
observed to date. It is very broad, rounded, and smooth (Fig.\ 1;
Smith et al.\ 2007). It looks like the light curve of SN~1987A
multiplied by a factor of $\sim$250. Can this shape be explained by
the photon diffusion model?

In generating a simple analytic light curve, we borrow directly from
the early study of Falk \& Arnett (1973, 1977), who studied this
physical situation. In general, the peak due to photon diffusion
becomes more luminous, more stretched-out in time, and smoother with
progressively larger initial radii. Following the scaling arguments
above, we assume a constant density envelope of 10 M$_{\odot}$ and
initial radius $R$=160 AU, with \td = \te = 70 days.

The SN will appear suddenly at $t = t_0$, when the blast wave reaches
the surface of the shell, delayed by a few weeks after the actual
explosion occurs because during that time the blast wave was expanding
through the opaque shell.  The initial rise is basically that of a
constant-velocity expanding blackbody, so that $L\propto t^2$ at early
times (Fig.\ 1). It levels off near maximum as the photosphere recedes
through the cooling ejecta, where we have adopted expressions from
equations (40) and (44) of Arnett (1982).  For a reasonable emitting
photosphere temperature of $T_{eff}$=10$^4$ K, the radius required at
the light curve peak on day 70 to account for the luminosity of
$L$=50$\times$10$^9$ L$_{\odot}$ is $R$=320 AU. Working backwards with
a constant expansion speed of 4000 km s$^{-1}$ gives an initial radius
of 160 AU at $t_0$. This agrees with our derived value for $R$ above.
 
The decline in luminosity following the peak at day 70 is determined
by thermal diffusion. The shape of the light curve depends on the
adopted values of \td \ and \te.  Arnett (1996) provides a highly
simplified model in which the SN light is due to diffusion of internal
radiation through a homogeneous expanding sphere.  In this case, the
light curve decline is expressed as
 
\begin{equation} 
L = L_{\rm max} \ e^{-[(\tau_{exp}t + t^2/2) / \tau_{exp} \tau_{diff}]} 
\end{equation} 

\noindent where $t$ is the time in days after $t_0$. Adopting \td =
\te = 70 days gives the decline rate shown in Figure 1, providing a
satisfactory fit to light curve of SN~2006gy until about day 160.
With a faster expansion speed and lower envelope mass, and hence,
smaller values for both \ \te \ and \ \td, the light curve would
evolve more rapidly.  Recently, Quimby et al.\ (2007) advocated just
such a model for SN~2005ap, which had an extremely high peak
luminosity exceeding that of SN~2006gy, but faded much faster.

The agreement of the early light curve of SN~2006gy with our simple
analytic light curve in Figure 1 is surely an oversimplification of
the actual situation.  The density profile resulting from the impact
of the SN debris with an opaque shell will not be a homogeneous
sphere, and the internal impact may continue over a timescale
comparable to or greater than \td.

The total radiated energy of the thermal diffusion model in Figure 1
is $E_{rad}$=1.4$\times$10$^{51}$ ergs, and the kinetic energy for
$\sim$10 M$_{\odot}$ expanding at 4000 km s$^{-1}$ is
$E_{kin}$=1.6$\times$10$^{51}$ ergs. It is mildly comforting that
these two are nearly equal, since that is about the highest efficiency
we can expect. However, this is likely an underestimate of the kinetic
energy of the SN because it is the mass of the accelerated pre-SN
envelope (see below), and does not include the mass of the stellar
ejecta themselves, which may be comparable. Thus, the {\it minimum}
total energy involved in SN~2006gy may be $\ga$3$\times$10$^{51}$
ergs, requiring a highly energetic explosion in any case.

\section{CAN THE RADIUS BE THAT BIG?} 
 
The requirements of the shell-shocked model as an explanation for the
early light curve of SN~2006gy are roughly the following: 1) the
initial radius of the ``star'' at shock breakout must have been
roughly 160 AU; 2) the radius of the photosphere at peak $\sim$70 days
after that must have been about twice as large, or $\sim$320 AU; 3)
the mass of the extended envelope needs to have been about 10
M$_{\odot}$.

Of course, it is nonsense to invoke a progenitor star with an actual
radius of 160 AU, since even the most extreme tenuously-bound red
supergiants have radii 20 times less (e.g., VY CMa; Smith et al.\
2001). So the essential question then is {\it how could the progenitor
star have had such a huge initial radius?}

The most likely answer is that it was not the star's true hydrostatic
radius, but rather, the photospheric radius of an opaque circumstellar
envelope. Smith et al.\ (2007) suggested that the progenitor of
SN~2006gy may have suffered an explosive mass-loss event in the decade
before the SN.  This would have been analogous to events observed in
very massive progenitor stars like $\eta$ Carinae, which ejected 10-15
M$_{\odot}$ in about a decade (Smith \& Owocki 2006; Smith et al.\
2003). A shell of that mass is sufficient to convert 10$^{51}$ ergs
into light.  High resolution spectra of SN~2006gy suggest that the
expansion speed for pre-SN mass loss was of order 200 km s$^{-1}$
(Smith et al.\ 2007). A mass of 10 M$_{\odot}$ ejected at those speeds
would require $\sim$4$\times$10$^{48}$ ergs, a substantial energy
release but still much less than that of a SN.

In our shell-shocked model, the luminosity in the main emission peak
of SN~2006gy comes from thermal diffusion of shock energy deposited
into an optically thick $\sim$10 M$_{\odot}$ extended circumstellar
envelope. Thus, the mass inferred from the diffusion timescale is the
envelope mass, not the stellar mass.

This brings up a critical distinction between our shell-shocked
diffusion model and previous CSM-interaction models for Type~IIn
SNe. In an optically thin interaction model (see Chevalier \& Fransson
1994), such as that invoked to explain SN~1994W by Chugai et al.\
(2004), a shock plows into a dense but optically thin CSM. Radiation
emitted directly from the swept-up shell of post-shock gas accounts
for the continuum luminosity of the SN, which can be considerably
greater than that of normal SNe~II. For SN~2006gy, Ofek et al.\ (2007)
suggeted a model where a Type~Ia explosion expands into a dense H-rich
CSM, and the continuous conversion of kinetic energy into radiation
powers the visual luminosity; this is similar to the models proposed
for SN~2002ic and 2005gj by Hamuy et al.\ (2003) and Aldering et al.\
(2006), respectively.

This type of CSM-interaction model cannot account for SN~2006gy
because {\it the extremely high CSM density required to power the
radiated luminosity will also cause that same circumstellar envelope
to be very opaque.} It is clear that a circumstellar mass of order 10
M$_{\odot}$ is required to reprocess 10$^{51}$ ergs of kinetic energy
into light (Smith et al.\ 2007; Ofek et al.\ 2007). If this much mass
is distributed inside a sphere with a radius corresponding to the
characteristic expansion radius of the SN ($R$=160 AU) then the
particle density is roughly $n$=2$\times$10$^{11}$ cm$^{-2}$ and the
optical depth $\tau$=$n \sigma R$ is greater than 300. This
automatically forces the situation back into the thermal diffusion
approximation described in this paper, because the radiation generated
in the CSM interaction cannot escape freely.  For such a large optical
depth, even relatively hard ($\sim$10 keV) X-rays will be thermalized.
The only way to avert this situation would be if the pre-shock CSM is
fully recombined so that the opacity is lower than for Thomson
scattering, allowing the massive envelope to be optically thin at
visual wavelengths. Given the high radiative luminosity involved, this
is unlikely, especially since high resolution spectra of the H$\alpha$
line (Smith et al.\ 2007) show that the pre-shock CSM is in fact
ionized.

In principle, then, the circumstellar matter ejected by the star
before the SN event is so dense and opaque that it mimics a gigantic
red supergiant envelope with a radius of more than 100 AU, although it
is not bound to the star and is probably expanding at 200 km s$^{-1}$
(Smith et al.\ 2007).  Thermal energy from the shock is deposited
throughout this very large envelope, so that it can escape without
suffering great adiabatic losses.  This fact accounts for the
exceptional luminosity of the SN when the shock breaks through the
circumstellar envelope.

Of course, after the shock escapes the pseudo-photosphere of the
opaque shell, it will propagate into the lower-density wind that must
surely reside outside that photosphere. It is this interaction that
likely gives rise to the characteristic Type~IIn spectrum of
SN~2006gy, while the optical continuum is generated by the thermal
radiation escaping from the shocked opaque envelope.  This model may
have the potential to explain many observed aspects of SN~2006gy that
have been grounds to favor the pair-instability SN hypothesis over CSM
interaction (Smith et al.\ 2007), such as the faint X-ray emission,
the weak pre-shock H$\alpha$ emission, the lack of any deceleration in
the shock, and the properties of the unusual broad P~Cygni absorption
feature in H$\alpha$.

\section{SUMMARY AND A PREDICTION} 
 
The diffusion model for SN~2006gy that we present here does not
require any new physical processes beyond those that account for
core-collapse SNe.  The only new ingredient is the hypothesis that the
SN progenitor must have ejected an opaque shell having
$\sim$10~M$_{\odot}$ extending to $\sim$160 AU shortly before it
exploded.  According to our model, the light curve of SN~2006gy
observed up to about day 170 comes entirely from energy deposited by
the initial blast and does not require any source of radioactive
energy or ongoing CSM interaction.

However, this internal energy source cannot last.  As the shocked
shell continues to expand, the radiative diffusion time becomes
shorter than the time since explosion, after which the light curve
will faithfully track the energy deposition rate.  Most of the kinetic
energy of the SN must reside in debris having velocity $\ga$4000 km
s$^{-1}$, and thermal energy deposition from the impact of this debris
with the expanding shell must terminate in $\la$1 yr.  After that,
there are only two other obvious sources of energy deposition to
account for the light curve.

The first is the decay of $^{56}$Co, which would produce an
exponential tail to the light curve having luminosity $L
\approx$1.4$\times$10$^{43}$ M($^{56}$Ni) $\exp(t/113.6 d)$ ergs
s$^{-1}$, where M($^{56}$Ni) is the mass of newly synthesized
$^{56}$Ni produced by the SN.  If this is the explanation for the
excess luminosity above the diffusion model after day 170, then about
8~M$_{\odot}$ of $^{56}$Ni would be required, indicated by the dashed
line in Figure 1.  On the other hand, failure to detect a continued
exponential tail in the visual light curve at late times does not
necessarily give an upper limit to M($^{56}$Ni) produced in SN~2006gy.
The bolometric flux may shift to near-infrared wavelengths as ejecta
cool, and it is quite possible that dust will form in the shocked
shell, reprocessing the luminosity into far infrared radiation.  This
cooling may also cause changes in the bolometric correction as the SN
evolves with time, and may play a role in the tail after day $\sim$160
seen in Figure 1.

A second source could be continued shock interaction with the
transparent CSM external to the opaque shell.  We see clear signs of
that interaction in the emission lines seen in the optical spectrum
and in the faint X-ray emission (Smith et al.\ 2007).  CSM interaction
could produce a lower-luminosity tail to the light curve of SN~2006gy
as long as the CSM density remains high.

\acknowledgments \scriptsize 
 
We acknowledge fruitful interaction with Roger Chevalier who prompted
us to consider in more detail the shell-shocked model of an energetic
supernova exploding in an expanded stellar envelope. We also thank S.\
Woosley, C.\ Wheeler, A.\ Filippenko, and S.\ Kulkarni for relevant
discussions.  N.S.\ benefits from continued collaboration and
interaction with all members of the U.C.\ Berkeley supernova group.


\end{document}